\documentclass[prl,twocolumn,showpacs,preprintnumbers,amsmath,amssymb]{revtex4}

\usepackage{epsfig}
\usepackage{graphicx}
\usepackage{dcolumn}
\usepackage{bm}

\newcommand{\vf}{v_{\rm F}}

\newcommand{\ef}{\epsilon_{\rm F}}
\newcommand{\tsci}{\tau_{\rm elastic}^{-1}}

\newcommand{\lB}{\ell_{\rm B}}
\newcommand{\sdc}{\sigma_{\rm dc}}

\begin{document}

\title{Electronic Properties of Two-Dimensional Carbon}
\author{N. M. R. Peres$^{1,2}$,  
F. Guinea$^{1,3}$, and A. H. Castro Neto
$^2$}

\affiliation{$^1$Department of Physics, Boston University, 590 
Commonwealth Avenue, Boston, MA 02215,USA}
\affiliation{$^2$Center of Physics and Department  of Physics,
Universidade do Minho, P-4710-057, Braga, Portugal}
\affiliation{$^3$Instituto de Ciencia de Materiales de Madrid, CSIC,
 Cantoblanco E28049 Madrid, Spain}

\begin{abstract}
 We present a theoretical description of the electronic properties of
 graphene in the presence of disorder, electron-electron interactions,
 and particle-hole symmetry breaking.
 We show that while particle-hole asymmetry, long-range Coulomb interactions, 
 and extended defects lead to the phenomenon of
 self-doping, local defects determine the transport and spectroscopic properties.
 Our results explain recent experiments in graphitic devices and predict
 new electronic behavior.
\end{abstract}

\maketitle

Carbon with sp$^2$ bonding has many allotropic forms
with different dimensionality: three-dimensional graphite \cite{BCP88}, 
one-dimensional nanotubes \cite{nanotubes}, zero-dimensional fullerenes
\cite{buckyball}, and two-dimensional Carbon, also known as graphene.
Graphene can be considered the {\it materia prima}
for the other forms of Carbon that can be obtained from it 
either by stacking (graphite),
wrapping (nanotubes), or creation of topological defects (fullerenes).
Hence, the electronic response of many Carbon based systems depends fundamentally
on the basic physics of graphene. 
Recent experiments in graphene-based devices have shown that it is possible 
to control their electrical properties, such as carrier type (particle
or hole), by the application of external gate voltage \cite{Netal04,outros}.
These experiments not only have opened doors to carbon-based
nano-electronics but also pose new questions on the
nature of the electronic properties in these systems. 

Being two-dimensional, true long range positional order of the
Carbon atoms is not possible at any finite temperature,
since thermal fluctuations introduce topological lattice defects such as 
dislocations (the Hohenberg-Mermim-Wagner theorem). 
Furthermore, because of the particular structure of the honeycomb lattice,
the dynamics of lattice defects in graphene belongs to
the generic class of kinetically constrained models \cite{DS00,RS03}, 
where defects are never completely annealed since their 
number is only weakly dependent on the annealing time \cite{DS00}. 
Indeed, defects are ubiquitous in carbon allotropes with sp$^2$ coordination, 
as confirmed by recent experiments \cite{Hetal04b}. 
Furthermore, lattice defects induced by proton
bombardment have been correlated to the appearance of magnetism in graphitic
samples \cite{magnet}, indicating the interplay between electron-electron
interactions and disorder.

Besides dislocations, graphene can sustain other types of extended
defects such as disclinations, edges, and micro-cracks. It is
known that certain edges (such as the zig-zag edge) 
lead to the appearance of localized states at the Fermi level 
\cite{WS00,Vetal05}.  Other defects such as pentagons and heptagons
(lattice disclinations) also admit localized states \cite{GGV92}.  
Furthermore, tight-binding calculations show that a combination of a
five-fold and seven-fold ring (a lattice dislocation) or a Stone-Wales defect 
(made up of two pentagons and two heptagons) also lead to a finite density of 
states at the Fermi level\cite{CEL96,MA01,DSL04}. The presence of
states at the Fermi level generated by defects 
can be tracked down to the particular
electronic structure of graphene. Each $\pi$-orbital contributes with
one electron (a half-filled band) and
the low-energy physics is described by a two-dimensional Dirac equation
with a vanishing electronic density of states at the Fermi level. The
vanishing of the density of states has two very important consequences:
the enhancement of the electron-electron interactions because of the
absence of electronic screening, and the formation of defect states
at the Fermi level. In this respect, our work complements the extensive literature on defects on
electronic systems described by the two dimensional Dirac
equation\cite{hirsch}. 

In the present work, we show that unscreened Coulomb interactions,
particle-hole symmetry breaking and defects play a fundamental role
in the electronic properties of graphene. In particular, graphene 
presents the phenomenon of {\it self-doping}, that is, charge 
transfer between extended defects and the bulk. In this case, 
depending on particle-hole symmetry breaking, graphene samples
can either have electron or hole pockets.  While self-doping
derives from extended defects and controls the number and type of
charge carriers, transport properties also
depend on point-like defects such as vacancies, and 
surface ad-atoms.  As we show in what follows, magneto-transport properties,
such as Shubnikov-de Hass oscillations and
quantum Hall effect (QHE), as well as spectroscopic quantities, such
as quasiparticle lifetimes and infrared reflectivity, can be
completely explained within this framework. Besides explaining published
experimental data, we also make new experimental predictions
that can be used as tests of our theory.

The bulk Hamiltonian of graphene can be written as:
${\cal H} = {\cal H}_0 + {\cal H}_V + {\cal H}_U$, where, 
\begin{equation}
{\cal H}_0 = - t \sum_{\sigma ; \langle i,j \rangle} c^\dag_{i,\sigma} c_{j, \sigma} + t'
\sum_{\sigma ; \langle \langle i,j \rangle \rangle } c^\dag_{i,\sigma} c_{j, \sigma} + h. c.
\label{hamil} 
\end{equation}
where $c_{i,\sigma}$ ($c_{i,\sigma}^{\dag}$) annihilates (creates) electrons
at the site ${\bf R}_i$ with spin $\sigma$ ($\sigma=\uparrow,\downarrow$), 
$t$ and $t'$ are the nearest neighbor and next-nearest
neighbor hopping energies, respectively. For ${\bf k} \to 0$ the electronic dispersion
is  $\epsilon_{\bf k} \approx 3 t' + v_{\rm F} | {\bf k} | +
(9 t' a^2 |{\bf k}|^2) / 4$. Notice that $t'$ does not change the structure of the
extended electronic wavefunctions, 
that remain described by the two-dimensional Dirac equation with
a Fermi-Dirac velocity $v_{\rm  F}  = (3 t a)/2$ ($a$ is the lattice spacing). 
${\cal H}_V$ describes the disorder,
\begin{eqnarray}
{\cal H}_V = \sum_{j} V_j n_j \, ,
\label{disorder}
\end{eqnarray} 
where $V_j$ is the potential strength (for vacancies, $V_j \to \infty$ at the
vacant site and is zero, otherwise), and $n_j =  \sum_{\sigma} c^\dag_{j,\sigma}
c_{j, \sigma}$. The electron interactions read,
\begin{eqnarray}
{\cal H}_U = (e^2/\epsilon_0) \sum_{i,j \, (i \neq j)} n_i n_j/|{\bf R}_i-{\bf R}_j| \, ,
\label{interaction}
\end{eqnarray}
$e$ is the electric charge, and $\epsilon_0$ is the dielectric constant. 

\begin{figure}[htf]
\begin{center}
\includegraphics[width=9cm]{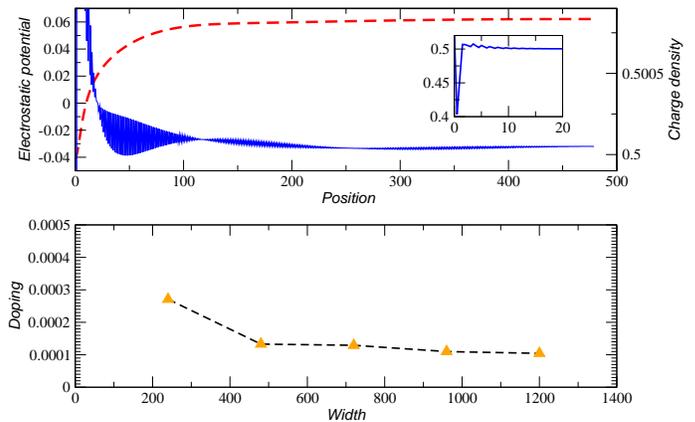}
\end{center}
\caption{
\label{self_doping}
Top:  Electrostatic potential (dashed line) and electronic charge (continuous
line) as a function of position
along a long graphene ribbon terminating on a zig-zag edge. 
Energies are given in units $t$, and distances in units of $a$. 
We have used $t' = 0.2 t$, and $e^2 / ( \epsilon_0 t a ) = 0.5$.
The  inset shows the details of the electronic density near the edge. Bottom:
Total additional charge per C atom in extended states as function of 
the ribbon width, $W$.}
\end{figure}

Standard derivations of localized states near defects assume electron-hole
symmetry, that is, $t'=0$. In this case, the localized states lie exactly at the Fermi level. 
In order to ensure charge neutrality,
these states are half filled, and there is no charge transfer between extended
and localized states. Electron-hole symmetry is broken by terms in the
Hamiltonian that allow for hopping between sites in the same sublattice, that
is, when $t' \neq 0$. In this case the band of localized states
near an extended defect acquires a bandwidth of order $t'$ and 
is shifted from Fermi energy.
Using (\ref{hamil}) and (\ref{interaction}), we have studied in the
Hartree approximation the induced charge transfer
between the edge and the bulk in a long graphene ribbon of width $W$.
In Fig.[\ref{self_doping}] we present the results for the induced
electrostatic potential, charge density, and charge transferred from a
zig-zag edge to (delocalized) bulk states. We have studied zig-zag edges
as long as $0.1 \mu$ m in graphene ribbons with up to $10^5$ atoms.
As one can clearly see in Fig.[\ref{self_doping}], self-doping effects are
suppressed as the width of the edge increases. The Coulomb energy
at the edge induced by a constant doping $\delta$ per Carbon atom 
is approximately $ \sim (\delta e^2/a) (W/a)$.
The charge transfer is arrested when the potential 
shifts the band of localized states to the Fermi energy,
that is, when $t' \approx (e^2/a) (W/a) \delta$. Hence the
self-doping is $\delta \sim ( t' a^2 ) / ( e^2 W )$.
The numerical results are consistent with this estimate. Note that the sign
of the induced charge depends on the sign of $t'$. 
The extended defects lead to a scattering
rate, $\tsci \sim \vf / W$.  The pocket induced by the self-doping
has a Fermi energy of 
$\ef \sim ( \hbar \vf \sqrt{\delta}) / a \sim t \sqrt{( t' a^2 ) / ( e^2  W)}$ 
and thus, $\hbar \tsci \ll \ef$ in wide ribbons. 
Hence, disorder due to extended defects does not 
smooth out the details of the small Fermi surface induced by the self-doping. 
For $t = 2.7$ eV \cite{BCP88}, $t' =0.2 t \approx 0.54$ eV (see below), 
and $a=1.42 \AA$, we get $\delta \sim 10^{-5} -10^{-4}$ per Carbon atom in a system with 
domains of typical dimensions $W \sim 0.1 - 1 \, \, \mu$ m, consistent with
experiments \cite{Netal04}. 

\begin{figure}[htf]
\includegraphics[width=7cm,angle=-90]{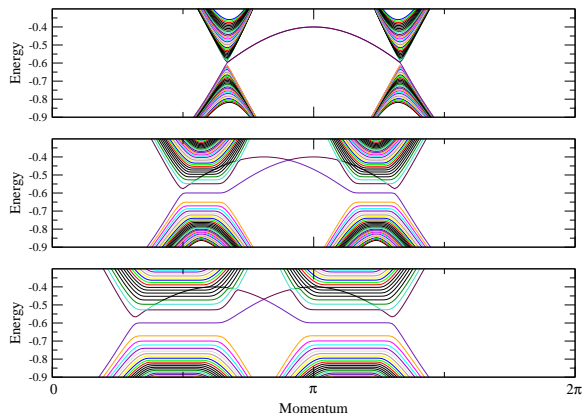}
\caption{
\label{ribbon}
Energy levels of a graphene ribbon of width 600 $a$ with
zig-zag edges as function of the momentum parallel to the edges. 
Energy in units of $t$, and $t'=0.2t$. Momentum in units of $1/ (2 \sqrt{3} a)$. 
Top: zero magnetic field; Centre: a magnetic flux per C atom of 
$\Phi= 0.00025 \, \Phi_0$; Bottom: $\Phi = 0.0005 \, \Phi_0$.}
\end{figure}

We can use the ribbon geometry described above to analyze the properties of
graphene in the integer QHE. 
A magnetic field $B$ changes the phases of the hopping terms in the
Hamiltonian (\ref{hamil}), leading 
to the appearance of bands of degenerate Landau levels. In the continuum
limit the Landau levels have energy 
$\epsilon_n = \pm v_{\rm F} l_{\rm B}^{-1} \sqrt{n}$, where $n$ is a positive
integer, $l_{\rm B} = \sqrt{\Phi_0 / B}$ is the cyclotron radius, and 
$\Phi_0 =h/e$ is flux quanta.  The Hall conductance can be obtained from
the number of bulk states that cross the Fermi level due to the presence
of an edge \cite{bert}. The calculated energy levels of a graphene ribbon 
in a magnetic field are shown in Fig.[\ref{ribbon}].
Notice that the momentum along the ribbon fixes the
position of the levels. When the position approaches the edges, the positive
energy levels rise in energy as for electron-like Landau levels in the 2D
electron gas, while the negative energy levels behave in a hole-like
fashion. There is a zero energy mode that splits into a set of electron and hole-like levels. Finally, the localized states at the edges are quite insensitive
to the applied magnetic field. If we fix the chemical potential above (below)
the zero mode energy, all bulk electron-like (hole-like) levels of lower energy give 
rise to crossings, and contribute to the Hall conductance. 
Each Landau level is doubly degenerated since the honeycomb lattice gives
rise to two inequivalent sets of Dirac fermions \cite{schakel}. Hence, the Hall conductance
arising from the bulk modes is: 
\begin{eqnarray}
\sigma_{{\rm Hall}} = 2 (2 N +1)  e^2/h \label{Hall} \, ,
\end{eqnarray}
and thus the Hall conductivity is quantized in odd number of $2 e^2/h$ (the
factor of $2$ comes from the spin degeneracy since cyclotron energy, 
$\hbar \omega_c = \sqrt{2} v_{\rm F} \hbar/l_{\rm B}$, is much larger than the Zeeman
energy, $g \mu_B B$, where $g \approx 2$ and $\mu_B$ the Bohr magneton). 
This analysis neglects corrections due to the additional band of localized
states at the edges induced by the structure of the ribbon. 

\hspace{1.5cm}

\begin{figure}[htf]
\begin{center}
\includegraphics*[width=8cm]{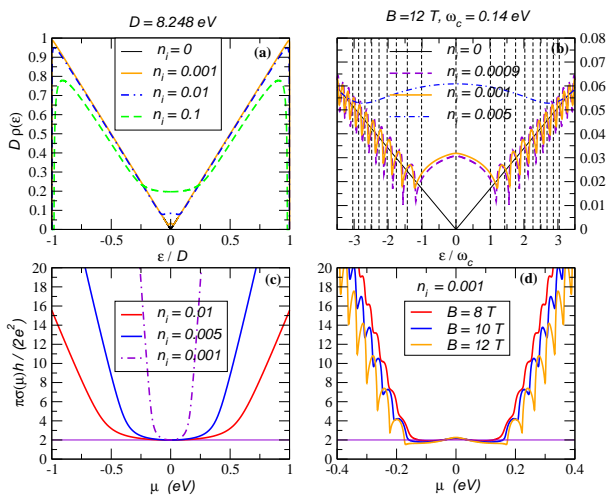}
\end{center}
\caption{
\label{dos} {\bf (a)}: Density of states as function of vacancy
concentration ($D$ is a high energy cut-off); {\bf (b)}:
As in (a), in the presence of an applied field. Dashed lines mark the
positions of the Landau levels in the absence of disorder; 
{\bf (c)} and {\bf (d)}: $\sigma(\mu)$ as function of the Fermi energy $\mu$.
$t'=0$ in all cases.}
\end{figure}

\begin{figure}[htb]
\begin{center}
\includegraphics[width=6cm]{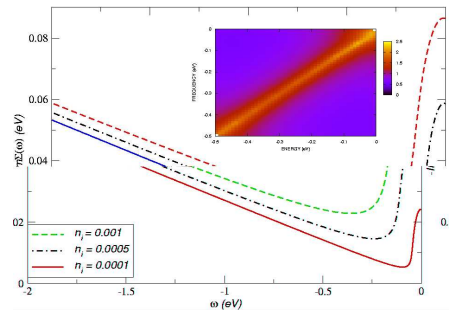}
\end{center}
\caption{\label{lifetime}$|\Sigma''(\omega)|$, as a function of energy
$\omega$. Inset: $A({\bf k},\omega)$. 
}
\end{figure}

While extended defects induce self-doping they do not change
significantly the electronic properties. However, point
defects such as vacancies or ad-atoms can change the transport 
properties even if they do not lead to self-doping. It is
known that weaker types of disorder do not change the semi-metallic nature of
a system described by the Dirac equation \cite{hirsch}.  
Nevertheless, point defects in the unitary limit can lead to a sharp
resonance at the Fermi energy \cite{hirsch}.  
We describe the effect of vacancies in the transport properties
using the Coherent Phase Approximation (CPA), that 
gives a good description of the  spectral and
transport properties of graphene,
except perhaps within a small region near Fermi energy where electronic
localization becomes important \cite{fisher}. Within CPA, the
one-particle spectral function is written as:
\begin{equation}
A ( {\bf k} , \omega ) = {\rm Im} \left[ 
\omega - \Sigma_{\rm CPA} ( \omega ) - 
\epsilon_{\bf k} \right]^{-1} 
\label{CPA} 
\end{equation} 
where electron self-energy, $\Sigma_{\rm CPA} ( \omega )$, has to be determined self-consistently
from (\ref{hamil}) and (\ref{disorder}). The electronic density of states,
$\rho(\omega) = \sum_{{\bf k}} A({\bf k},\omega)$, 
as function of applied magnetic field and impurity concentration, in the continuum
limit, is shown in
Fig.[\ref{dos}](a)-(b). Unlike for weaker forms of disorder, vacancies induce a
finite density of states at the Fermi level. Furthermore, ${\rm Im}
\Sigma_{\rm CPA} ( \omega )$ is finite at the Fermi level, and is monotonically
{\it increasing} as $\omega \to 0$ indicating that vacancies have a strong effect
on the Dirac fermions.  This function can be considered an intrinsic linewidth 
as measured in ARPES. Besides the effect of disorder, the lifetime of quasiparticles 
has also an intrinsic contribution from interaction effects associated with eq.~(\ref{interaction})
giving a contribution 
$e^2 / v_{\rm F} \, | \omega |$\cite{GGV96}, which is monotonically {\it decreasing}
as $\omega \to 0$. Therefore, the final result, as shown in Fig.[\ref{lifetime}], predicts that 
that the total linewidth of quasiparticles shows a minimum.

The frequency dependent conductivity can be written as:
\begin{equation}
\sigma(\omega,T) = 2e^2/(\hbar \pi)
\int_{- \infty}^\infty d \epsilon K (\epsilon,\omega)
[f(\epsilon,T)-f(\epsilon+\omega,T)]/\omega 
\label{conductivity_dc} 
\end{equation}
where $K (\epsilon,\omega)$ is a conductivity kernel,
and $f(\epsilon,T)$ is the Fermi-Dirac distribution
function. At low temperatures, the d.c. conductivity, 
$\sigma(\mu)=\sigma(\omega=0,T)$, 
is approximately proportional to $K(\mu,\omega=0)$, and can be experimentally
measured by the application of a gate voltage, $V$, to a graphene plane 
(in the presence of a voltage $V$ the chemical potential shifts from $\mu$
to $\mu+e V$). In Fig. [\ref{dos}] (c)-(d)
we show $\sigma(\mu)$.  A non-conventional
$\mu$ dependence is found when compared to other 2D electronic systems,
in agreement with experiments \cite{Netal04}.  As the temperature rises, the
d.c. conductivity is found to increase with
temperature, as the thermally excited carriers contribute (as in the
case of a narrow gap semiconductor) and is also in agreement with the 
experimental data \cite{Netal04}. When $\mu+e V=0$ the low
temperature conductance per plane has a universal value, independent of
disorder and magnetic field \cite{F86}, $\sdc =  4 e^2  / ( \pi h )$. A similar
universal behavior was predicted for d-wave superconductors \cite{L93}.
Because of particle-hole symmetry breaking, $t' \neq 0$, 
we expect an asymmetry in the
Shubnikov-de Haas oscillations as function of gate voltage.
We can quantify this asymmetry by looking at Landau levels $j^*$ and $-j^*-1$ with
$j^* \gg 1$. It is possible to show \cite{next} that: 
\begin{equation}
\left|t'/t \right| \approx \sqrt{2}/12 (j^*+3/2)^{-1} (j^*+1)^{-1/2} (\lB/a) \,.
\end{equation}
For $j^* \approx 6$ for $B = 12$ T ($\lB = 52$ \AA) \cite{Netal04},
we find $| t' / t | \approx 0.2$, which is the value used in this work.

\begin{figure}[htf]
//
\begin{center}
\includegraphics[width=7cm]{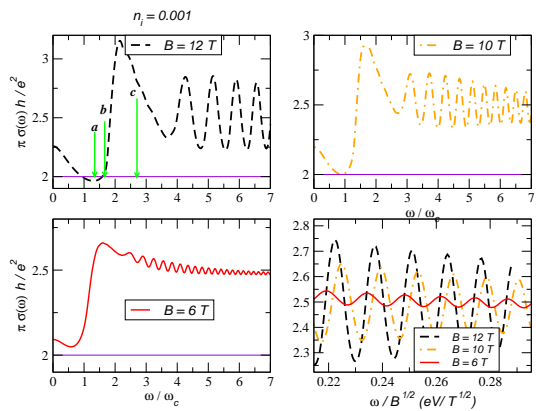}
\end{center}
\caption{
\label{sigma_W_T_B} $\sigma(\omega)$ 
 for different magnetic fields and $n_i =0.001$, 
using the parameters given in the text. 
The lower right graph shows a scaling of the curves as
 function of $\omega/\sqrt{B}$.}
\end{figure}

In Fig.[\ref{sigma_W_T_B}] we show $\sigma(\omega)$ in the presence
of impurities and finite magnetic field. 
The existence oscillations in the conductivity at low temperatures is
clearly seen. An analysis of the transitions to the different peaks shows that 
those involving the $j=0$ Landau level  are suppressed and that the
energy of the lowest
Landau levels is significantly shifted by the disorder.

We have analyzed the influence of lattice defects on the electronic 
properties of graphene. Our results show that: 
({\it i}) Extended defects lead to the existence of localized states, or resonances, near the
Fermi level of the pure system. In the absence of electron-hole symmetry,
these states lead to self-doping, and to the existence of electron or hole Fermi
pockets with $10^{-4} - 10^{-5}$ electrons per unit cell for domains of sizes $0.1 - 1 \mu$m. 
({\it ii}) An integer QHE with a quantized Hall conductivity in odd values
of $2 e^2/h$. 
({\it iii}) Point defects lead to a finite inverse scattering time in undoped
graphene, which decreases with increasing energy. As the inelastic
contribution rises linearly with temperature, a minimum in the lifetime
observed in ARPES experiments is predicted.
({\it iv}) The d.c. conductivity in undoped graphene has a universal value at low
doping and temperatures. In undoped samples, it rises with temperature, as
the number of thermally activated carriers increases. 
({\it v}) The optical conductivity in a magnetic field is modulated
due to Landau levels.
 
N. M. R. P. and F. G. acknowledge 
the Quantum Condensed Matter Visitor's Program at Boston University for
support. N. M. R. P. acknowledges Funda\c{c}\~ao para
a Ci\^encia e Tecnologia for a sabbatical grant.
A. H. C. N. was supported by the NSF grant DMR-0343790.
We thank D. N. Basov, A. Geim, G.-H. Gweon, A. Lanzara, and S. W. Tsai for 
illuminating discussions. 




\begin{thebibliography}{99}

\bibitem{BCP88}
N. B. Brandt, S. M. Chudinov, and Ya. G. Ponomarev, 
in {\it Modern Problems in Condensed Matter Sciences}, 
V. M. Agranovich and A. A. Maradudin, eds.
(North Holand, Amsterdam, 1988).

\bibitem{nanotubes}
S. G. Rao {\it et al.}
 Nature {\bf 425}, 36 (2004).

\bibitem{buckyball}
M. Sawamura {\it et al.}, Nature {\bf 419}, 702 (2002).


\bibitem{Netal04}
K. S. Novoselov {\it et al.}, 
Science {\bf 306}, 666 (2004);
 K. S. Novoselov {\it et al.}, 
cond-mat/0503533; 
A. K. Geim {\it et al.}, unpublished.

\bibitem{outros}
C.~Berger {\it et al.}, cond-mat/0410240; 
Y.~Zhang {\it et al.}, cond-mat/0410314; 
Y.~Zhang {\it et al.}, cond-mat/0410315.





\bibitem{DS00}
L. Davison, and D. Sherrington, 
J. Phys. A {\bf 33}, 8615 (2000).

\bibitem{RS03}
F. Ritort, and P. Sollich,
Adv. in Phys. {\bf 52}, 219 (2003).



\bibitem{Hetal04b}
A. Hashimoto {\it et al.}, Nature {\bf 430}, 870 (2004).

\bibitem{magnet}
P. Esquinazi {\it et al.}, 
Phys. Rev. Lett. {\bf 91}, 227201 (2003).

\bibitem{WS00}
K. Wakayabashi, and M. Sigrist, 
Phys.Rev.Lett. {\bf 84}, 3390 (2000).
K. Wakayabashi,
Phys. Rev.B {\bf 64}, 125428 (2001).

\bibitem{Vetal05}
M- A. H. Vozmediano, {\it et al}, cond-mat/0505557.

\bibitem{GGV92}
J. Gonz\'alez {\it et al.}, 
Phys. Rev. Lett. {\bf 69}, 172 (1992); {\it ibid},
Phys.Rev.B {\bf 63}, 134421 (2001).

\bibitem{CEL96}
J.C. Charlier {\it et al.}, 
Phys.Rev.B {\bf 53}, 11108 (1996).

\bibitem{MA01}
H. Matsumura and T. Ando, 
J. Phys. Soc. Japan {\bf 70}, 2657 (2001).

\bibitem{DSL04}
E. J. Duplock {\it et al.}, 
Phys. Rev. Lett. {\bf 92}, 225502 (2004).

\bibitem{hirsch}
See, for instance, P. J. Hirschfeld, and W. A. Atkinson,
J. Low. Temp. Phys. {\bf 126}, 881 (2002).

\bibitem{bert}
B. I. Halperin, Phys. Rev. B {\bf 25}, 2185 (1982).

\bibitem{schakel}
For a single Dirac fermion, see, A. M. J. Schakel, Phys. Rev. D {\bf 43}, 1428 (1991).

\bibitem{GGV96}
J. Gonz\'alez {\it et al.}, 
Phys. Rev. Lett. {\bf 77}, 3589 (1996).

\bibitem{fisher}
T. Senthil and M. P. A. Fisher,  
 Phys. Rev. B {\bf 60}, 6893 (1999). 

\bibitem{F86}
E. Fradkin,  Phys. Rev. B {\bf 33}, 3257 (1986).

\bibitem{L93}
P. A. Lee, Phys. Rev. Lett. {\bf 71}, 1887 (1993).

\bibitem{next}
N. M. R. Peres {\it et al.}, unpublished.


\end{thebibliography}
\end{document}